\documentclass[12pt]{article}
\pdfoutput=1
\usepackage{amsmath,amssymb,amsfonts}
\usepackage{a4wide,epsfig,psfrag,scalefnt}
\usepackage[dvipsnames]{xcolor}
\usepackage{braket}
\usepackage{placeins}
\usepackage{breqn}
\usepackage{slashed}
\usepackage{enumitem}
\usepackage[numbers,sort&compress]{natbib}
\usepackage{caption}
\usepackage{subcaption}

\usepackage{comment}

\graphicspath{ {figs/} }

\allowdisplaybreaks

\newcommand{\ep}{\epsilon}

\newcommand{\lm}{l_x}

\begin{document}

\title{\vskip-3cm{\baselineskip14pt
    \begin{flushleft}
     \normalsize CERN-TH-2023-148, P3H-23-059, TTP23-035, ZU-TH 56/23
    \end{flushleft}} \vskip1.5cm
  Three-loop $b\to s\gamma$ vertex with current-current operators
  }

\author{
  Matteo Fael$^{a}$,
  Fabian Lange$^{b,c}$,
  Kay Sch\"onwald$^{d}$,
  Matthias Steinhauser$^{b}$
  \\
  {\small\it (a) Theoretical Physics Department, CERN,}\\
  {\small\it 1211 Geneva, Switzerland}
  \\
  {\small\it (b) Institut f{\"u}r Theoretische Teilchenphysik,
    Karlsruhe Institute of Technology (KIT),}\\
  {\small\it 76128 Karlsruhe, Germany}
  \\
  {\small\it (c) Institut f{\"u}r Astroteilchenphysik,
  Karlsruhe Institute of Technology (KIT),}\\
  {\small\it 76344 Eggenstein-Leopoldshafen, Germany}
  \\
  {\small\it (d) Physik-Institut, Universit\"at Z\"urich, Winterthurerstrasse 190,}\\
  {\small\it 8057 Z\"urich, Switzerland}
}

\date{}

\maketitle

\thispagestyle{empty}

\begin{abstract}

\noindent
We compute three-loop vertex corrections to $b\to s\gamma$ induced by
current-current operators. The results are presented as
expansions in $m_c/m_b$ with numerical coefficients
which allow to cover all relevant values for the heavy quark masses
in different renormalization schemes.
Moreover we provide for the first time analytic results for the 
next-to-leading order contribution.
Our results present an important building block to the
next-to-next-to-leading order interference contributions
of the current-current operators $Q_1$ and $Q_2$
with the electric dipole operator $Q_7$.

\vspace*{5em}

\end{abstract}

\thispagestyle{empty}

%- }}}

\thispagestyle{empty}

\newpage

%- {{{ Introduction:

\section{\label{sec::intro}Introduction}

The inclusive rare decay $ B \to X_s \gamma$, where $X_s$ is any
charmless hadronic state 
of strangeness $S=-1$, is an important probe to look for phenomena beyond the
Standard Model. It is thus of prime importance to both measure it with highest
accuracy, and to provide precise predictions. 

The current combination of various experimental
measurements~\cite{CLEO:2001gsa,BaBar:2007yhb,Belle:2009nth,BaBar:2012fqh,BaBar:2012eja,Belle:2014nmp}
leads to the CP- and isospin-averaged branching
ratio~\cite{HeavyFlavorAveragingGroup:2022wzx} 
\begin{eqnarray}
   {\cal B}(B\to X_s \gamma)\Big|_{E_\gamma>1.6~\mbox{GeV}} &=& (3.49\pm
   0.19)\times 10^{-4}\,,         
\end{eqnarray}
where $E_0=1.6$~GeV is a lower cut on the energy of the
photon. The current uncertainty of about 5\% is expected to be reduced by
a factor two once the full Belle-II data set is
analysed~\cite{Belle-II:2018jsg}.

The most precise theory prediction from
Refs.~\cite{Misiak:2015xwa,Czakon:2015exa} also has an uncertainty of
5\% and is given by
\begin{eqnarray}
   {\cal B}(B\to X_s \gamma)\Big|_{E_\gamma>1.6~\mbox{GeV}} &=& (3.40\pm
   0.17)\times 10^{-4}\,,         
        \label{eq::bsg_th}
\end{eqnarray}
where the updates from Ref.~\cite{Misiak:2020vlo} are taken into account.  

The decay $B \to X_s \gamma$ is well approximated by the 
decay of a free quark $b \to X_s \gamma$. Perturbative QCD corrections
are calculated within the framework of an effective theory obtained after 
decoupling the top quark, the $Z$, $W$ and Higgs bosons.
The weak effective theory then contains four-quark and 
dipole-type operators (see e.g.\ Refs.~\cite{Buchalla:1995vs,Chetyrkin:1997gb} and Section~\ref{sec::tech}).
The prediction in Eq.~(\ref{eq::bsg_th}) includes next-to-next-to-leading order (NNLO)
QCD corrections. However, for the contribution arising from the
interference between the four-quark operators ($Q_1$ and $Q_2$) and the
electromagnetic dipole operator ($Q_7$) only an interpolation from a large
charm quark mass~\cite{Misiak:2006ab,Misiak:2010sk} to a massless charm
quark~\cite{Czakon:2015exa} is available. This is responsible for 3\% of the
uncertainty cited in Eq.~(\ref{eq::bsg_th}).  
The remaining theoretical uncertainties come from 
unknown higher-order corrections (3\%) and the input and
non-perturbative parameters (2.5\%).
In this paper we provide important contributions that are necessary to eliminate 
the uncertainty due to the charm mass interpolation.

\begin{figure}[t]
  \centering
  \begin{tabular}{cc}
    \includegraphics[width=0.45\textwidth, trim = 36 468 36 36]{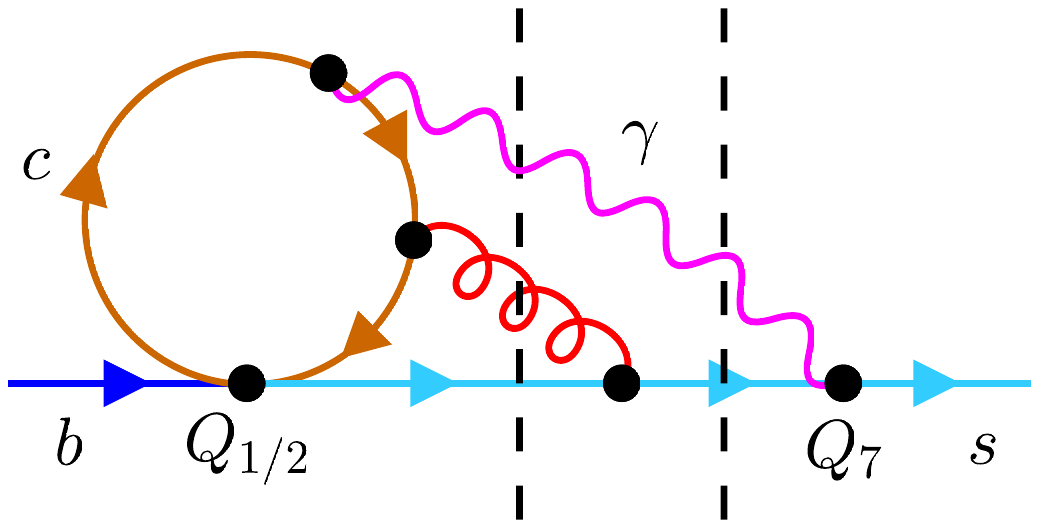}
    &
    \includegraphics[width=0.45\textwidth, trim = 36 468 36 36]{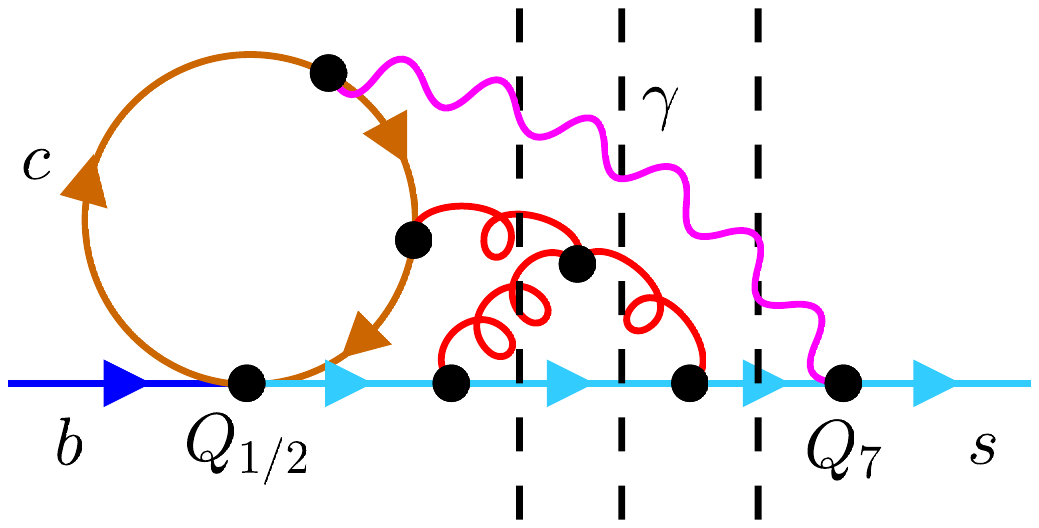}
  \end{tabular}
  \caption{\label{fig::o2o7_prop}Three- and four-loop sample diagrams for the
  interference of $Q_{1,2}$ and $Q_7$. The dashed black lines represent possible cuts
  through the diagrams.}
\end{figure}

In order to calculate the interference of $Q_{1,2}$ and $Q_7$, one often
applies the method of reverse unitarity~\cite{Anastasiou:2002yz} to map the
calculation of the interference into the evaluation of cuts of two-point
functions.  Such contributions occur for the first time at next-to-leading
order (NLO) where three-loop diagrams, as the one on the left of
Fig.~\ref{fig::o2o7_prop}, have to be considered. These diagrams have both
two- and three-particle cuts as indicated by the dashed
lines. Correspondingly, at next-to-next-to-leading order (NNLO) one has to
consider four-loop diagrams which in general have two-, three- and
four-particle cuts.

In this paper we concentrate on the two-particle cut contribution.  It can be
obtained by calculating QCD corrections to the $b\to s \gamma$ vertex and
subsequently performing the integration over the two-particle phase space.  Of
course, in the physical decay rate all cuts going through the photon line
have to be considered.  In fact, individual contributions contain divergences
which cancel only when all cuts are taken together and after including the
counterterm from renormalization of the ultraviolet divergences. The latter is
conveniently done for the complete contributions and available from
Ref.~\cite{Misiak:2017woa} which is why we only provide results for the bare
amplitude.

There are a number of NNLO results available for the interference of $Q_{1,2}$
and $Q_7$. Analytic results for the light-fermion two-particle cut
contributions are available from Ref.~\cite{Bieri:2003ue} in an expansion
for $m_c/m_b\to 0$.  The corresponding four-particle cut terms are available
from Ref.~\cite{Ligeti:1999ea}. The contributions with closed charm and
bottom quark loops have been computed in
Refs.~\cite{Boughezal:2007ny,Misiak:2020vlo}. A numerical approach has been
used to solve the differential equations for the master integrals with
boundary conditions computed for large values of $m_c$.  The same approach has
been used in Ref.~\cite{Misiak:2017woa} to extend the NLO results by one order
in $\epsilon$ and to obtain all counterterm contributions. The non-fermionic
contribution has been computed in the large-$m_c$ limit in
Refs.~\cite{Misiak:2006ab,Misiak:2010sk} and for massless charm quarks in
Ref.~\cite{Czakon:2015exa}. These results are used in
Refs.~\cite{Misiak:2015xwa,Czakon:2015exa} to construct an interpolation which
induces the 3\% uncertainty mentioned above.

Recently, Ref.~\cite{Greub:2023msv} has considered the subclass of diagrams
for the $b \to s \gamma$ vertex where no bottom quark propagator is present in
the loop. Here we provide an independent check of these results and also
compute the more involved contributions with internal bottom quarks.

The paper is organized as follows. In Section~\ref{sec::tech} we
introduce the notation and present the setup of our calculation. We
then discuss in Section \ref{sec::MIs} the calculation of the master
integrals at two and three loops.  Results are presented in
Section~\ref{sec::res} and we conclude in
Section~\ref{sec::conclusions}.  In Appendix~\ref{app::2loop} we provide further
details on the analytic calculation of the two-loop
master integrals and in Appendix~\ref{app::ab} analytic expressions for the
individual NLO contributions are given.

%- }}}

%- {{{ Technicalities:

\section{\label{sec::tech}Technicalities}

The relevant weak interaction Lagrangian for the calculation
of radiative $B$ meson decays is given by a linear combination of four-quark
and dipole-type operators and can be written as
\begin{eqnarray}
  {\mathcal L}_{\rm weak} = \frac{4G_F}{\sqrt{2}} V_{ts}^\star V_{tb}
   \sum_i C_i (\mu) Q_i\, ,
\end{eqnarray}
where the three operators considered in this paper are given by
\begin{eqnarray}
Q_1  &=& (\bar{s}_L \gamma_{\mu} T^a c_L) (\bar{c}_L     \gamma^{\mu} T^a b_L)\,,\nonumber\\
Q_2  &=& (\bar{s}_L \gamma_{\mu}     c_L) (\bar{c}_L     \gamma^{\mu}     b_L)\,,\nonumber\\
Q_7  &=&  \frac{em_b}{16\pi^2} (\bar{s}_L \sigma^{\mu \nu}     b_R) F_{\mu \nu}.
\label{eq::operators}
\end{eqnarray}
Here, $s$, $c$ and $b$ are the fields of the strange, charm and bottom quarks,
respectively.  The subscripts $L$ and $R$ denote the projection to left- and
right-handed states and $T^a=\lambda^a/2$, where $\lambda^a$ are the Gell-Mann
matrices. $F^{\mu\nu}$ is the field strength tensor of the photon field, $e$
is the electric charge, $m_b$ the bottom quark mass, and
$\sigma^{\mu\nu} = \mathrm{i}[\gamma^\mu,\gamma^\nu]/2$.

We can write the amplitude for $b\to s\gamma$ as
\begin{eqnarray}
  A &=& { \frac{4G_F m_b^2}{\sqrt{2}} V_{ts}^\star V_{tb}  } \, M^\mu \varepsilon_\mu\,,
        \label{eq::A}
\end{eqnarray}
where $\varepsilon^\mu$ is the polarization vector of the photon.
We parameterize $M^\mu$ as
\begin{eqnarray}
  M^{\mu} &=& \bar{u}_s(p_s)P_R\left(
                t_1 { \frac{q_\gamma^\mu}{m_b} }
                + t_2 { \frac{ {p_b^\mu} }{m_b} }
                + t_3 \gamma^\mu \right) u_b(p_b)
                \,,
                \label{eq::MAmu}
\end{eqnarray}
where all momenta are incoming and on-shell. We construct projectors
to extract the three scalar coefficients $t_1$, $t_2$ and $t_3$, which
depend on the ratio $m_c/m_b$.  The function $t_1$ does not contribute
to $b\to s\gamma$ since the corresponding tensor structure vanishes
once we contract $M^\mu$ with the photon polarization vector in Eq.~(\ref{eq::A}).  Furthermore, from
the Ward identity $q_\gamma^\mu M_\mu=0$ we have
\begin{eqnarray}
  t_3 &=& - { \frac{1}{2} t_2 }\,,
            \label{eq::WI}
\end{eqnarray}
which we use as a cross-check for our results.  Note that this
relation holds for renormalized quantities, i.e., in particular for
the two-loop amplitudes. Therefore all counterterm contributions
obtained from multiplicative renormalization fulfil the Ward identity.
However, the relation~\eqref{eq::WI} does not hold for the NNLO
contributions from the bottom quark mass counterterm since they
are not obtained from a global renormalization factor. As a
consequence, we do not expect that the bare three-loop amplitude
fulfils Eq.~\eqref{eq::WI} per se, but only in combination with the
bottom mass counterterm contributions.

\begin{figure}[t]
  \centering
  \begin{tabular}{ccc}
    \includegraphics[width=0.23\textwidth, trim = 36 342 36 36]{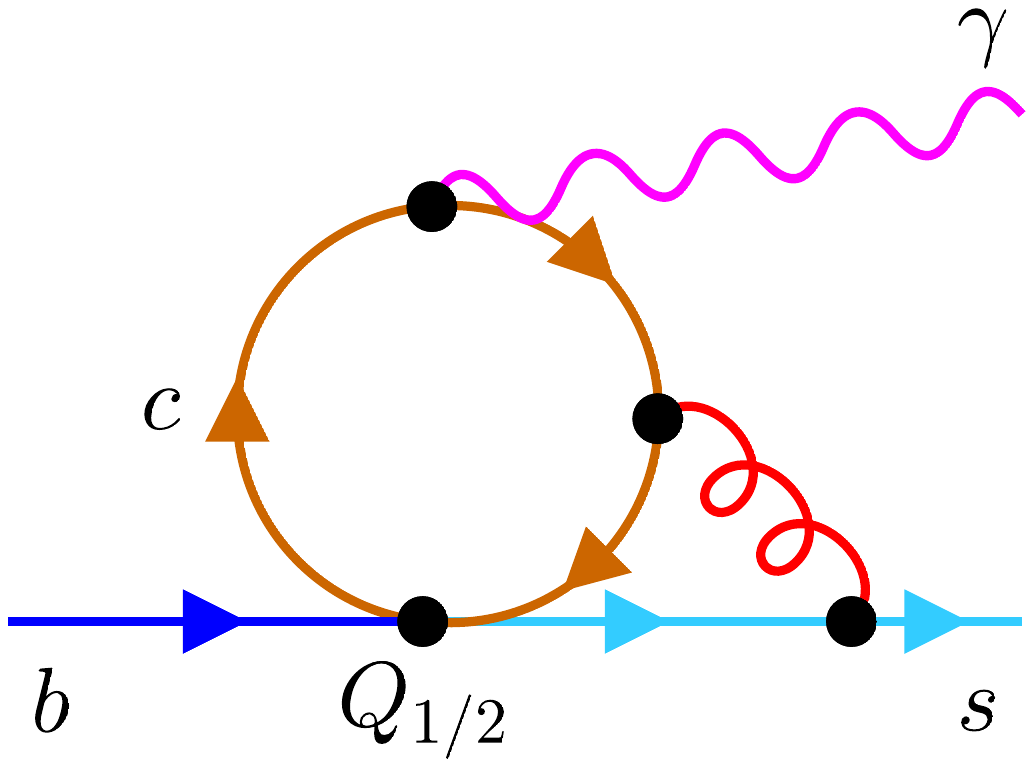}
    &
    \includegraphics[width=0.23\textwidth, trim = 36 360 36 36]{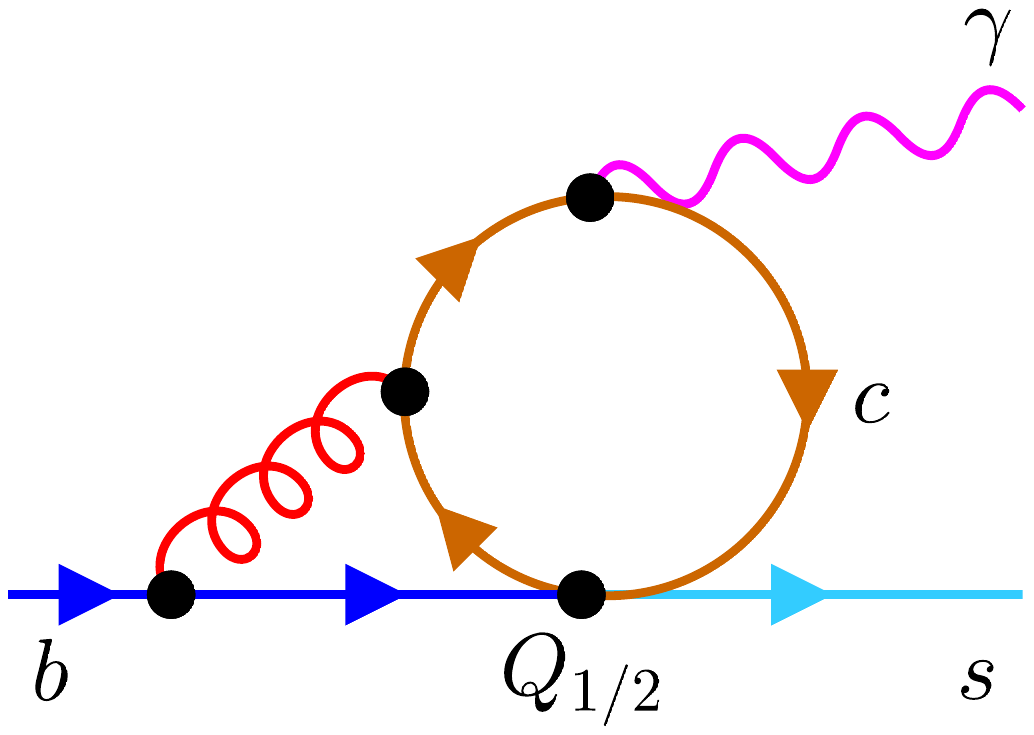}
    &
    \\
    \includegraphics[width=0.25\textwidth, trim = 36 390 36 36]{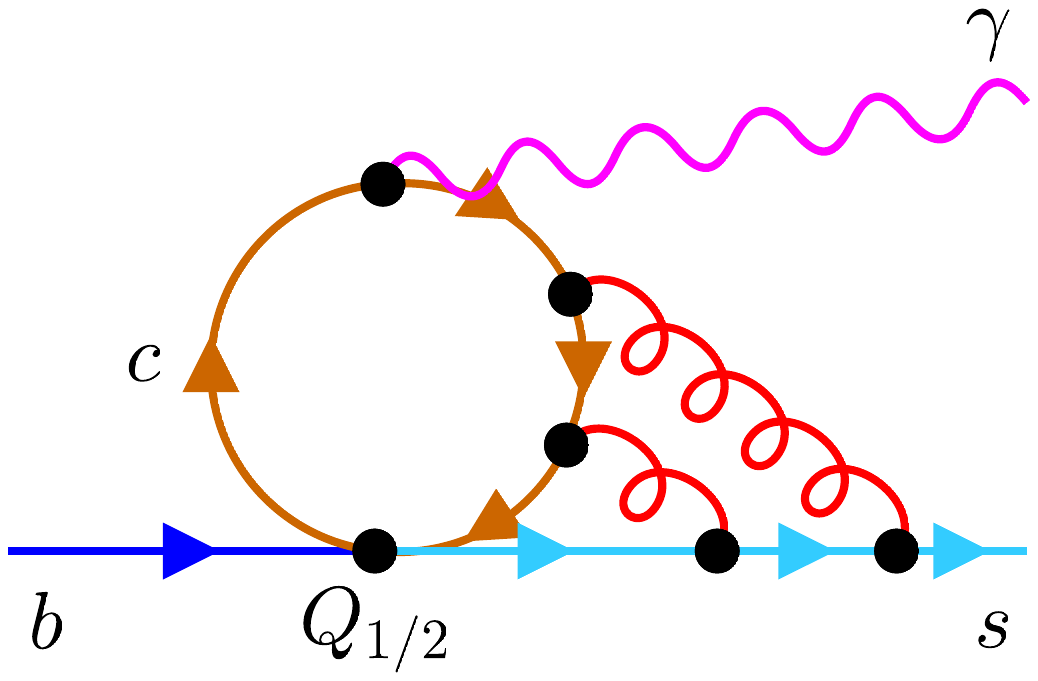}
    &
    \includegraphics[width=0.26\textwidth, trim = 36 417 36 36]{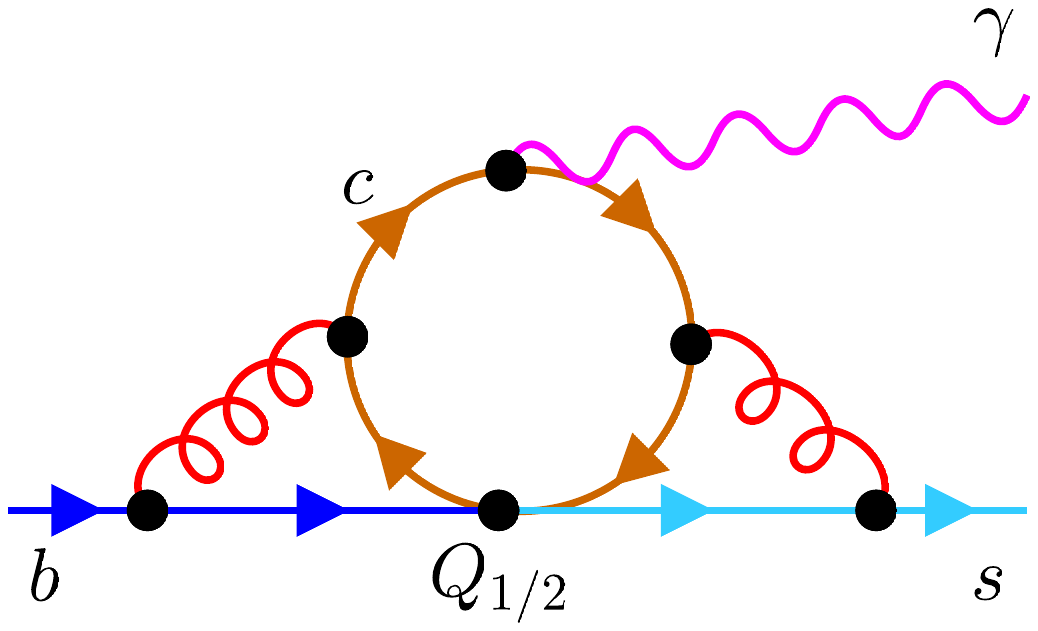}
    &
    \includegraphics[width=0.25\textwidth, trim = 36 393 36 36]{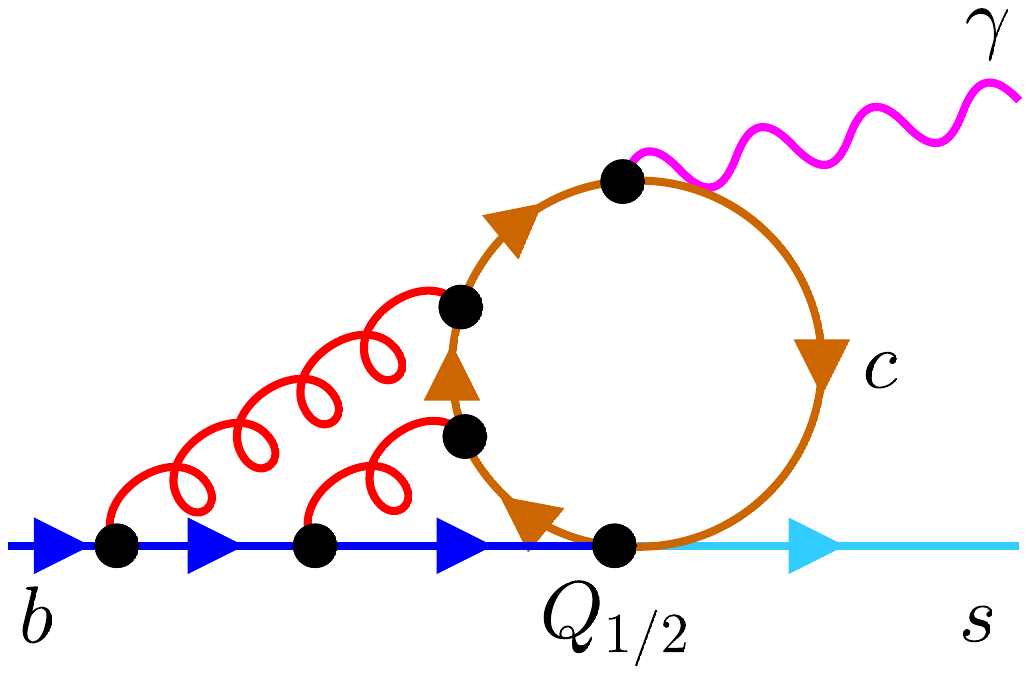}
    \\
    \includegraphics[width=0.27\textwidth, trim = 36 377 36 36]{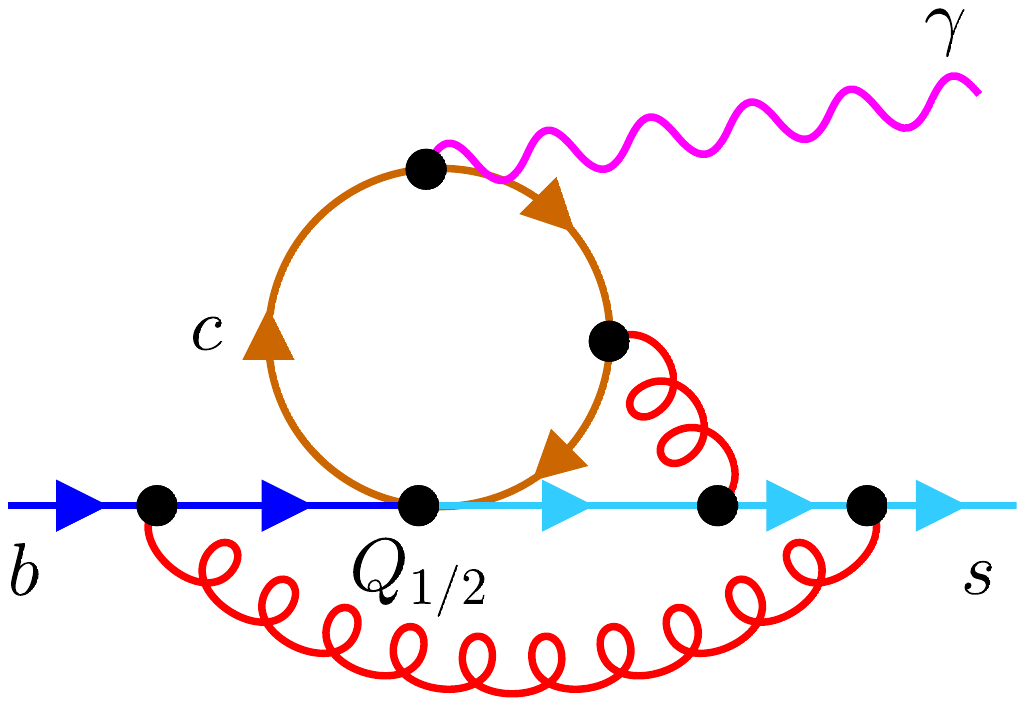}
    &
    \includegraphics[width=0.24\textwidth, trim = 36 376 36 36]{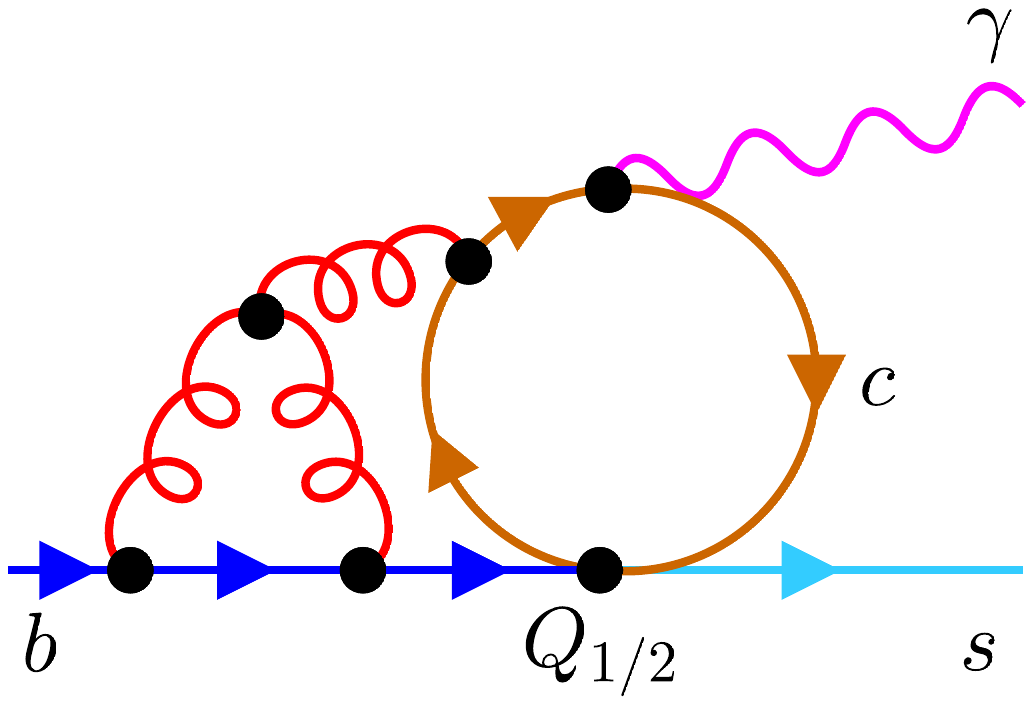}
    &
    \includegraphics[width=0.25\textwidth, trim = 36 391 36 36]{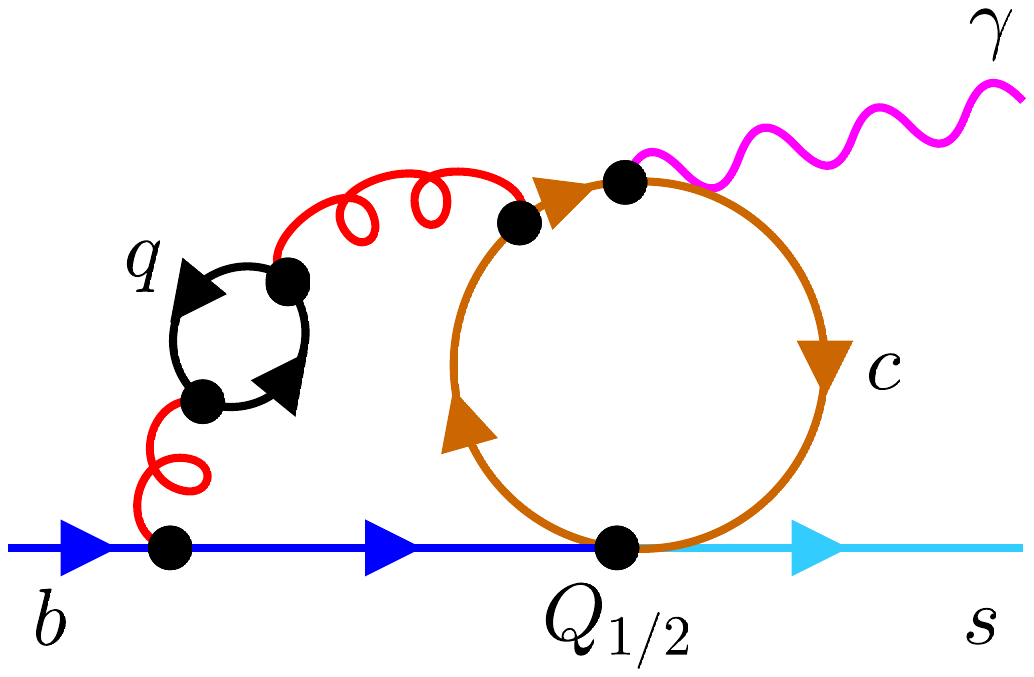}
    \\
    \includegraphics[width=0.30\textwidth, trim = 36 446 36 36]{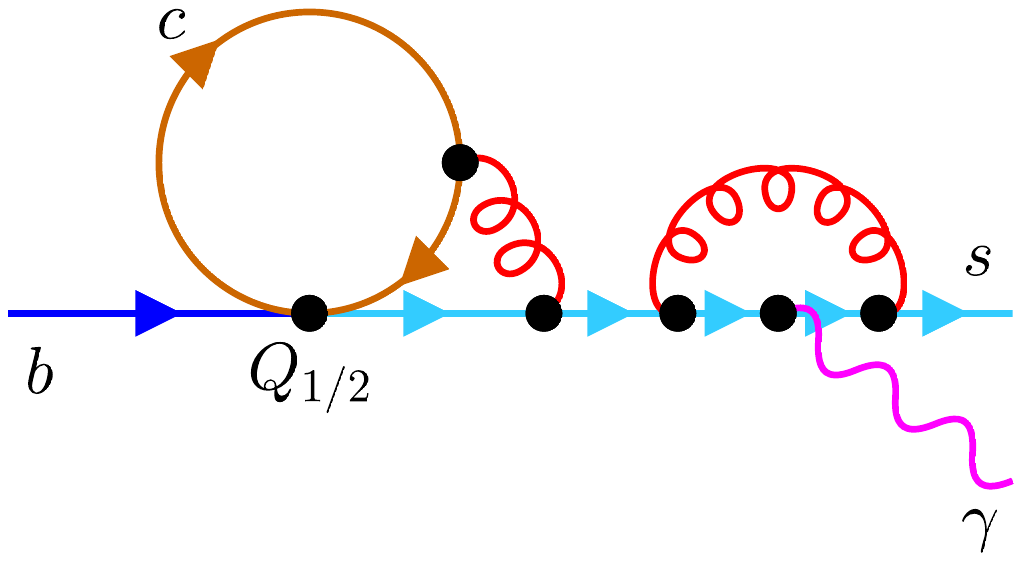}
    &
    \includegraphics[width=0.28\textwidth, trim = 36 412 36 36]{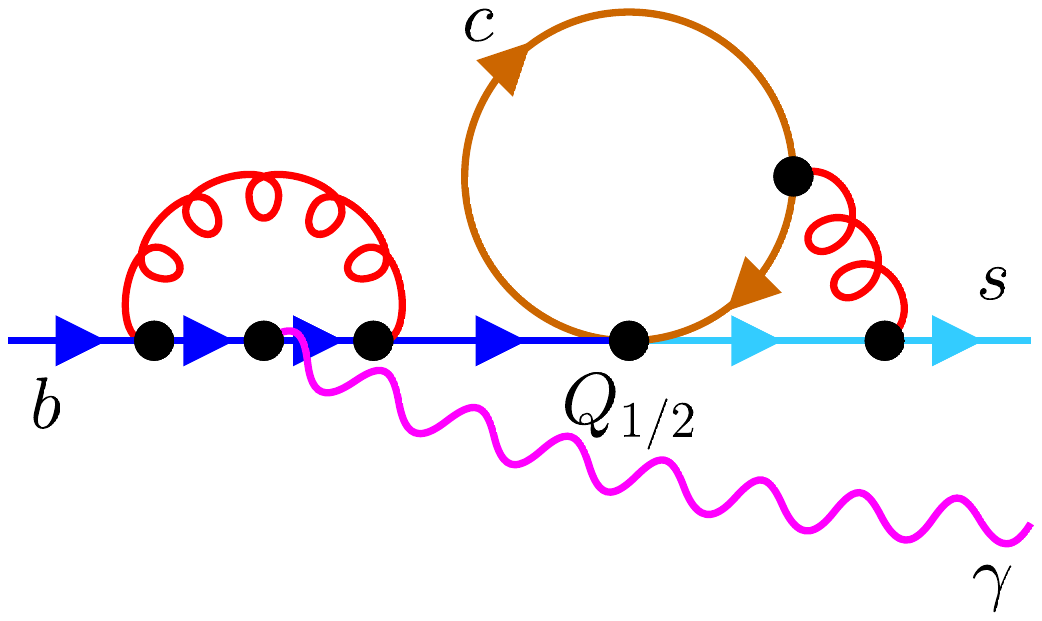}
    &
    \includegraphics[width=0.25\textwidth, trim = 36 373 36 36]{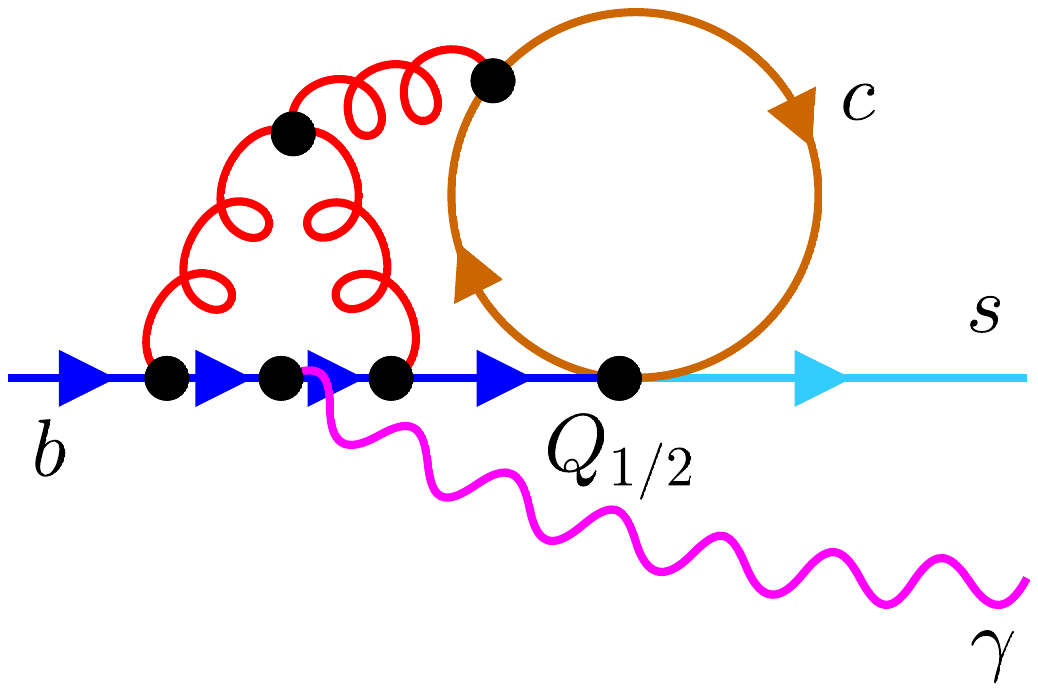}
  \end{tabular}
  \caption{\label{fig::diags}Two- and three-loop sample diagrams contributing
  to the decay vertex $b\to s\gamma$. No QCD corrections to external lines
  are considered.}
\end{figure}

Since the scalar coefficients $t_1$, $t_2$ and $t_3$ depend only on
the ratio $m_c/m_b$ and not on the external momenta, we can
straightforwardly obtain the two-particle cut contribution to the
decay rate:
\begin{eqnarray}
  \Gamma(b \to X_s^p \gamma)\Big|_{E_\gamma > E_0} ~=~ 
  \frac{G_F^2 m_b^5 \alpha_{\mathrm{em}}}{32 \pi^4} \left|V_{ts}^* V_{tb} \right|^2
  \sum_{i,j=1}^8 C_i(\mu_b) \, C_j(\mu_b) \, \hat{G}_{ij}\, ,
\end{eqnarray}
where the sum includes the operators in Eq.~(\ref{eq::operators}), the penguin
operators and the chromomagnetic dipole operators.
We write the perturbative expansion of the interference terms as
\begin{eqnarray}
\hat{G}_{ij} = \hat{G}^{(0)}_{ij} 
  + \frac{\alpha_s}{4\pi} \,  \hat{G}^{(1)}_{ij} 
  + \left( \frac{\alpha_s}{4\pi} \right)^2 \, \hat{G}^{(2)}_{ij} + \ldots
  \,.
\end{eqnarray}
In this paper we compute the two-particle cut contributions to $\hat{G}_{17}$
and $\hat{G}_{27}$ taking into account only tree-level contributions on the
$Q_7$ side. The only other non-zero contribution with two-loop corrections on
the $Q_{1,2}$ and one-loop corrections on the $Q_7$ side can be obtained from
lower-order results.  Sample diagrams for the three-loop corrections on the
$Q_{1,2}$ side are given in Fig.~\ref{fig::diags}.  $\hat{G}_{17}$ and
$\hat{G}_{27}$ are obtained by taking the interference between the
$b \to s \gamma$ amplitude in Eq.~\eqref{eq::MAmu} and the tree-level one
mediated by the operator $Q_7$.  The subsequent integration over the
$d$-dimensional two-particle phase space leads to (i=1,2)
\begin{eqnarray}
  \hat{G}_{i7}^{2P,Q_7^{\rm tree}} =
  { - } \, 
  \mathrm{Re} 
  \left[
  { \frac{t_2^{Q_i}}{2} } + \left(3-2\epsilon\right) t_3^{Q_i}
  \right]
  \, \frac{ e^{\gamma_E\epsilon} }{ {8}} \, \frac{\Gamma(1-\epsilon)}{\Gamma(2-2\epsilon)}
  \, ,
  \label{eq::comb_t2_t3}
\end{eqnarray}
where $\gamma_E$ is the Euler's constant. The superscripts on the
left-hand side indicate that this formula provides the two-particle cut
contributions where no loop corrections are considered for $Q_7$.
The last factor comes from the $d$-dimensional two-particle phase space.  Note
that in case we use the relation $t_3=-t_2/2$, the first factor becomes proportional to
$(1-\epsilon)$ and we recover Eq.~(3.5) of Ref.~\cite{Misiak:2017woa}.

For the computation of the $b \to s \gamma$ vertex at two and three
loops we use \texttt{qgraf}~\cite{Nogueira:1991ex} to generate all 30
and 591 diagrams, respectively; see Fig.~\ref{fig::diags} for
sample Feynman diagrams.  We process them with
\texttt{tapir}~\cite{Gerlach:2022qnc} and use
\texttt{exp}~\cite{Harlander:1998cmq,Seidensticker:1999bb} to prepare
\texttt{FORM}~\cite{Vermaseren:2000nd,Kuipers:2012rf,Ruijl:2017dtg}
code for their evaluation.  We then apply the projectors and perform
the Dirac and colour algebra~\cite{vanRitbergen:1998pn} to
express the form factors as linear combination of Feynman integrals
belonging to 10 and 181 integral families and two and three loops,
respectively.  The scalar integrals are reduced to master integrals
by employing
integration-by-parts~\cite{Tkachov:1981wb,Chetyrkin:1981qh} as well as
Lorentz-invariance relations~\cite{Gehrmann:1999as} and the Laporta
algorithm~\cite{Laporta:2000dsw} as implemented in
\texttt{Kira}~\cite{Maierhofer:2017gsa,Klappert:2020nbg}.  We use
\texttt{Fermat}~\cite{fermat} as back-end to process the coefficients.
Before the actual reduction we reduce sets of simpler sample integrals
for each integral family and feed them to an improved version of
\texttt{ImproveMasters.m}~\cite{Smirnov:2020quc} to find a good basis
of master integrals in which the denominators of the coefficients
factorize in the kinematic and the space-time
variable~\cite{Smirnov:2020quc,Usovitsch:2020jrk}.  The integrals in
the amplitude are then reduced to the good basis.  As last step we
exploit symmetry relations between the integral families with
\texttt{Kira} to reduce the number of master integrals at three
loops from $3975$ to $479$. At two loops we have 14 master
integrals.  We perform our calculation for general QCD gauge
parameter $\xi$ and check that $\xi$ drops out in the final result.

%- }}}

%- {{{ Master integral calculations at 2 and 3 loop:
\newpage
\section{\label{sec::MIs}Computation of the master integrals}

In this Section we describe the computation of the master integrals which
we encounter in the calculation of the $b \to s \gamma$ vertex at two-
and three-loop order. 
We use \texttt{LiteRed}~\cite{Lee:2012cn,Lee:2013mka} and a subsequent
reduction with \texttt{Kira} to establish differential equations in
\begin{eqnarray}
  x &=& \frac{m_c}{m_b}
  \label{eq::x}
\end{eqnarray}
for the master
integrals~\cite{Kotikov:1990kg,Kotikov:1991hm,Kotikov:1991pm,Remiddi:1997ny}.
At two-loop order we observe poles in the differential equations at
the physical cuts $x=0$ and $1/2$.
At three-loop order we have
additional spurious poles at various values of $x$.

At two loops we manage to obtain analytic expressions for all master
integrals appearing at NLO, and in the bottom mass counterterm at NNLO.
This extends the results of Ref.~\cite{Misiak:2017woa} where the
master integrals for the two- and three-particle cut contributions
have been obtained only by solving the differential equations
numerically with boundary conditions fixed in the large charm mass
limit.  Furthermore, analytic expansions around $m_c \to 0$ have been
presented including terms only up to order $(m_c^2/m_b^2)^3$.  Here,
instead, we provide for the master integrals contributing to the
two-particle cuts their analytic expressions with exact dependence on
$m_c/m_b$, which allows to obtain precise numerical results
for an arbitrary ratio of the heavy quark masses.

For the analytic calculation of the two-loop master integrals we use
the algorithm outlined in Ref.~\cite{Ablinger:2018zwz}.  We do not
transform the system of differential equations into a so-called
canonical or $\epsilon$ form (see
e.g.\ Refs.~\cite{Henn:2013pwa,Lee:2014ioa}), but decouple coupled
blocks of the differential system with the help of {\tt
  OreSys}~\cite{ORESYS} and {\tt Sigma}~\cite{Schneider:2007}. This
leads to a higher order differential equation for a single integral of
the block which can be solved via the factorization of the
differential operator as implemented in {\tt
  HarmonicSums}~\cite{HarmonicSums}.  The boundary conditions of the
integrals are provided in the limit $m_c \to \infty$ via a large mass
expansion. For the matching of the boundary conditions, it is
convenient to first solve the differential equations in the variable
$y = m_b/m_c$ since expansions for $y \to 0$ are easier to compute.
Afterwards we perform an analytic continuation to the variable $x
=m_c/m_b$ (further details are given in the Appendix).

To solve the integrals in the variable $y$ we introduce generalized
iterated integrals over the alphabet
\begin{align} 
    \frac{1}{y}, \frac{1}{1+y}, \frac{1}{1-y}, \frac{1}{2+y}, \frac{1}{2-y}, \sqrt{4-y^2}
    ~.
    \label{eq::alphabet}
\end{align}
Iterated integrals containing the first three letters yield Harmonic
Polylogarithms (HPLs)~\cite{Remiddi:1999ew}.  
A representation of the iterated
integrals in terms of rational letters is beneficial in order to
evaluate them to high precision with public programs like {\tt
  ginac}~\cite{Vollinga:2004sn}.  Thus we have to rationalize the
occurring square root valued letter with the variable change
\begin{align}
    x &= \frac{\sqrt{w}}{1+w} \,, \nonumber\\
    w &= \frac{1 - \sqrt{1-4{x^2}}}{1 + \sqrt{1-4{x^2}}} \,,
    \label{eq::w}
\end{align}
on the integrals which contain the new letters.  

For the HPLs with argument $x$ we do not apply the variable change.  
In case the new letters come together with the HPL letters, we have to introduce 
the additional letters $\sqrt{y}/(1 + y)$ and $\sqrt{y}/(1 - y)$ for individual
iterated integrals after changing from argument $x$ to $w$.
However, we see that contributions containing them
cancel in the final amplitude and we can express the results purely in terms
of HPLs with arguments $x$ and $w$, which enables a fast and
precise numerical evaluation for an arbitrary mass ratio.
Explicit analytic results for two-loop quantities originating from the
interference of the operators $Q_{1,2}$ and $Q_7$ can be found in
Section~\ref{sec::res_NLO} and are provided in electronic format in the 
supplementary material~\cite{progdata}.

For the master integrals at three loops we apply the ``expand and match'' approach developed in
Refs.~\cite{Fael:2021kyg,Fael:2022rgm,Fael:2022miw} to obtain a semi-analytic
expansion which covers all physically relevant values for $x$ that may arise 
when using different mass schemes for the charm and the bottom quark.  
The numerical values $m_c \sim 1~\mbox{GeV}, \ldots,
1.7~\mbox{GeV}$ and $m_b\sim 4.2~\mbox{GeV}, \ldots, 5.0~\mbox{GeV}$
correspond to $x\in[0.2,0.4]$.

The basic idea of this approach is to use the differential equations
in order to construct deep series expansions of all master integrals
around several values of $x_0$.  We make an ansatz for the master
integrals as Taylor expansion in $x-x_0$ (or a power-log expansion if
$x_0$ is a singular point) and insert the ansatz into the differential
equations. With \texttt{Kira} we solve the resulting system of linear
equations for the expansion coefficients and express them in terms of
a minimal set of initial values.

In our case we choose $x_0=1/5$ as a first expansion point.  The
boundary conditions are fixed by evaluating the master integrals
numerically with a precision of $60$ digits at this point with
\texttt{AMFlow}~\cite{Liu:2022chg}, which implements the
auxiliary-mass-flow
method~\cite{Liu:2017jxz,Liu:2020kpc,Liu:2021wks,Liu:2022tji,Liu:2022mfb}.
We use \texttt{Kira} with \texttt{Fermat} as back-end for the
reduction.  The precise {\tt AMFlow} results allow to fix all
coefficients of the expansions around $x_0 = 1/5$. We also perform an
{\tt AMFlow} run at $x=1/10$ to cross check the expansion constructed
around $x_0 = 1/5$.

This expansion alone already covers the physically relevant $x$ region
mentioned above.  In practical applications it is convenient to
work with an expansion around $x_0=0$. To obtain a good precision of
the expansion coefficients we introduce new expansions. First we
construct an expansion around $x_0=1/10$ that we match to the $x_0=1/5$
expansion at $x=0.15$.  In a second step we produce a power-log
expansion around $x_0=0$ that is matched at $x=1/15$ to the previous
one.

%- }}}

%- {{{ Two- and three-loop results:

\section{\label{sec::res}Two- and three-loop results}

In the following we present results for the bare NLO and NNLO
contributions.  Sample two- and three-loop Feynman diagrams are shown
in Fig.~\ref{fig::diags}.

%- {{{ NLO:

\subsection{\label{sec::res_NLO}NLO}

NLO corrections to $b\to s\gamma$ involving current-current operators
are known since long in the
literature~\cite{Ali:1990tj,Greub:1996jd,Greub:1996tg,Buras:2001mq,Buras:2002tp}.
In Ref.~\cite{Misiak:2017woa} $\hat{G}_{17}^{(1)}$ and
  $\hat{G}_{27}^{(1)}$ have been computed up to order $\epsilon$ and
the counterterm contributions to $\hat G^{(2)}_{ij}$ have been
extracted, see Eqs.~(4.1) to (4.4) of Ref~\cite{Misiak:2017woa}. For
convenience we repeat the relevant formulae which involve two-particle
(2P) cuts: 
\begin{eqnarray}
  \hat{G}_{27}^{(1)2\mathrm{P}} 
%&\equiv& \hat{G}_{27}^{(1)2P(d)} + \hat{G}_{27}^{(1)2P(u)} ~=~ 
  &=&
-\frac{92}{81\ep} + f_0(z) + \epsilon f_1(z) + {\mathcal O}(\ep^2)\,,
  \nonumber\\
\hat{G}_{27}^{(1)m,2\mathrm{P}} 
%&\equiv& \hat{G}_{27}^{(1)m,2P(d)} + \hat{G}_{27}^{(1)m,2P(u)} ~=~
  &=&
-\frac{1}{3\ep^2} + \frac{1}{\ep} r_{-1}(z) + r_0(z) + \epsilon r_1(z) + {\mathcal
                                O}(\ep^2)
                                \,,
\label{eq::G27}
\end{eqnarray}
with $z=m_c^2/m_b^2$. The superscript ``m'' denotes the
contribution from bottom quark renormalization. Furthermore, we have
\begin{eqnarray}
        \hat{G}_{17}^{(1)2\mathrm{P}}   &=& -\frac{1}{6} \hat{G}_{27}^{(1)2\mathrm{P}}\,,\nonumber\\
        \hat{G}_{17}^{(1)m,2\mathrm{P}} &=& -\frac{1}{6} \hat{G}_{27}^{(1)m,2\mathrm{P}}\,.
\end{eqnarray}
The function $f_0(z)$ which enters the NLO prediction
is usually written as
\begin{eqnarray}
  f_0(z) &=& - \frac{1942}{243} + 2 \mbox{Re} \left[ a(z) + b(z) \right]\,.
  \label{eq::f0_a_b}
\end{eqnarray}
For the funtions $a(z)$ and $b(z)$ there are no closed analytic expressions
known so far but only threefold integral
representations~\cite{Buras:2002tp}.

In this work we provide independent cross checks for $f_0(z)$, $f_1(z)$,
$r_{-1}(z)$, $r_0(z)$ and $r_1(z)$ and present for the first time their
analytic expressions in terms of HPLs.\footnote{We note that in the
  context of $b\to s \ell^+\ell^-$ analytic results for the two-loop form
  factors have been obtained in Refs.~\cite{deBoer:2017way,Asatrian:2019kbk}. However, the 
  limit from $q^2\not=0$ to $q^2=0$ is highly non-trivial such that
  the results cannot be used for $b\to s\gamma$.} Analytic results for $a(z)$ and
$b(z)$ can be found in Appendix~\ref{app::ab}.  For illustration we show the
expressions for $f_0$, $r_{-1}$ and $r_0$: 
\begin{align} 
    f_0 &= 
    C_F 
    \Biggl[
      -\frac{971+1916 w+1602 w^2+1916 w^3+971 w^4}{162 (1+w)^4}
      +\frac{2 w H_0(w)^3}{3 (1+w)^2}
      \nonumber \\ &
      +\frac{8 w \big(27+57 w+26 w^2+7 w^3+5 w^4\big) H_0(w)}{27 (1+w)^5}
      -\frac{16 w \big(2+3 w+2 w^2\big) H_0(x)^3}{9 (1+w)^4}
      \nonumber \\ &
      -\frac{2 w \big(-1-2 w+4 w^2+6 w^3+3 w^4\big) H_0(w)^2}{3 (1+w)^6}
      -\frac{8 w \big(1+w^2\big) H_{0,0,-1}(w)}{3 (1+w)^4}
      \nonumber \\ &
      -\frac{8 \big(5+29 w+54 w^2+29 w^3+5 w^4\big) H_{-1}(w)}{27 (1+w)^4}
      +\frac{16 w^2 \big(3+13 w+15 w^2+6 w^3\big) H_{0,-1}(w)}{9 (1+w)^6}
      \nonumber \\ &
      -\frac{16 w \big(1-\sqrt{w}+w\big)}{9 (1+w)^6} \big(3+8 w+8 w^2+3 w^3+2 \sqrt{w}+3 w^{3/2}+2 w^{5/2}\big) H_{1,0}(x)
      \nonumber \\ &
      +\frac{16 w \big(1+\sqrt{w}+w\big)}{9 (1+w)^6} \big(3+8 w+8 w^2+3 w^3-2 \sqrt{w}-3 w^{3/2}-2 w^{5/2}\big) H_{-1,0}(x)
      \nonumber \\ &
      +\frac{16 w \big(3+9 w+13 w^2+9 w^3+3 w^4\big) H_{-1,0}(w)}{9 (1+w)^6}
      -\frac{8 w \zeta_3}{(1+w)^2}
      \nonumber \\ &
      -\frac{32 w \big(3+9 w+13 w^2+9 w^3+3 w^4\big) H_{-1,-1}(w)}{9 (1+w)^6}
      -\frac{16 w \big(2+3 w+2 w^2\big) H_{0,1,0}(x)}{3 (1+w)^4}
      \nonumber \\ &
      +\frac{16 w \big(2+3 w+2 w^2\big) H_{0,-1,0}(x)}{3 (1+w)^4}
      -\frac{16 w \big(1+3 w+w^2\big) H_{-1,0,0}(w)}{3 (1+w)^4}
      \nonumber \\ &
      +\pi ^2 \biggl(
              -\frac{2 w}{27 (1+w)^6} \big(15+60 w+94 w^2+84 w^3+27 w^4-12 \sqrt{w}-36 w^{3/2}
              \nonumber \\ &
              -36 w^{5/2}-12 w^{7/2}\big)
              -\frac{2 w \big(3+8 w+3 w^2\big) H_0(w)}{3 (1+w)^4}
              +\frac{8 w \big(5+12 w+5 w^2\big) H_{-1}(w)}{9 (1+w)^4}
      \biggr)
    \Biggr]\,,
    \\ 
    r_{-1} &= - C_F \Biggl[ \frac{3}{4} + \frac{1}{27}\pi^2 + \frac{3}{2}z \Biggr] \,,
    \\ 
    r_0 &=
    C_F 
    \Biggl[
        \frac{315-386 w+315 w^2}{108 (1+w)^2}
        -\frac{64 w H_0(w)}{27 (1+w)^2}
        +\frac{\big(1+14 w+w^2\big) H_0(w)^3}{27 (1+w)^2}
        +\frac{4 w H_0(x)^4}{9 (1+w)^2}
        \nonumber \\ &
        +\frac{w \big(96+222 w-2 w^2-15 w^3-13 w^4\big) H_0(w)^2}{54 (1-w) (1+w)^4}
        -\frac{8 \big(1+38 w+w^2\big) H_0(x)^3}{27 (1+w)^2}
        \nonumber \\ &
        +\frac{128 w H_{-1}(w)}{27 (1+w)^2}
        +\frac{\big(13+124 w+348 w^2+124 w^3+13 w^4\big) H_{-1}(w)^2}{27 (1+w)^4}
        \nonumber \\ &
        -\frac{2 w \big(96+222 w-2 w^2-15 w^3-13 w^4\big) H_{0,-1}(w)}{27 (1-w) (1+w)^4}
        -\frac{8 \big(1+38 w+w^2\big) H_{0,1,0}(x)}{9 (1+w)^2}
        \nonumber \\ &
        -\frac{\big(13+124 w+348 w^2+124 w^3+13 w^4\big) H_{-1,0}(w)}{27 (1+w)^4}
        -\frac{4 \big(1+14 w+w^2\big) H_{0,0,-1}(w)}{9 (1+w)^2}
        \nonumber \\ &
        -\frac{2 \big(1-\sqrt{w}+w\big)}{27 (1+w)^4} \big(7+136 w+136 w^2+7 w^3+151 \sqrt{w}+365 w^{3/2}+151 w^{5/2}\big) H_{1,0}(x)
        \nonumber \\ &
        +\frac{2 \big(1+\sqrt{w}+w\big)}{27 (1+w)^4} \big(7+136 w+136 w^2+7 w^3-151 \sqrt{w}-365 w^{3/2}-151 w^{5/2}\big) H_{-1,0}(x)
        \nonumber \\ &
        +\frac{8 \big(1+38 w+w^2\big) H_{0,-1,0}(x)}{9 (1+w)^2}
        +\frac{16 w H_{0,0,1,0}(x)}{3 (1+w)^2}
        -\frac{16 w H_{0,0,-1,0}(x)}{3 (1+w)^2}
        \nonumber \\ &
        -\frac{2 \big(5+34 w+5 w^2\big) \zeta_3}{9 (1+w)^2}
        +\pi ^2 \Biggl(
                -\frac{1}{1296 (1-w) (1+w)^4} \big(161-573 w-2246 w^2+486 w^3
                \nonumber \\ &
                -75 w^4-57 w^5+6912 \sqrt{w}+11328 w^{3/2}-11328 w^{7/2}-6912 w^{9/2}\big)
                -\frac{20 w H_0(w)}{9 (1+w)^2}
                \nonumber \\ &
                +\frac{w H_0(w)^2}{9 (1+w)^2}
                +\frac{2 \big(1+74 w+w^2\big) H_{-1}(w)}{27 (1+w)^2}
                +\frac{4 w H_{-1}(w)^2}{9 (1+w)^2}
                -\frac{4 w H_{0,-1}(w)}{9 (1+w)^2}
                \nonumber \\ &
                -\frac{4 w H_{-1,0}(w)}{9 (1+w)^2}
        \Biggr)
        +\frac{2 \pi ^4 w}{45 (1+w)^2}
    \Biggr]
    ~,
    \end{align}
  where $x$ and $w$ are defined in Eqs.~(\ref{eq::x}) and~(\ref{eq::w}),
  respectively. These respresentations can be used for $x\le1/2$ since for
  $x>1/2$
  the HPLs develop imaginary parts.
  Note that the HPLs in $f_0$ and $r_0$ can be
  expressed in terms of classical polylogarithms.
  This is not the
  case for the ${\cal O}(\epsilon)$ terms $f_1$ and $r_1$.

The comparison with Eq.~(\ref{eq::f0_a_b}) allows us to extract the function
$\mbox{Re}(a(z)+b(z))$.  This is the first time that analytic results for
these functions are presented.  Before only a three-dimensional integral
representation~\cite{Buras:2002tp} and analytic expansions for
large~\cite{Misiak:2010sk} and small~\cite{Buras:2002tp} charm quark masses
were available.

Our analytic results for $r_0$
%(see Eq.~(\ref{eq::G27}))
and $r_1$ are expressed in terms
of HPLs up to weight~4 and~5, respectively, where as argument we
have both $x$ and $w$.  We find agreement with the analytic expansions given
in Refs.~\cite{Greub:1996jd,Misiak:2017woa} and can determine the constant $C$
entering $r_1$ in Eq.~(5.5) of Ref.~\cite{Misiak:2017woa}. We obtain
\begin{align}
    C &= -\frac{488}{9}-\frac{179 \pi ^2}{54}-\frac{80 \pi ^4}{27}+\frac{64 \zeta_{3}}{3}+\frac{16 \pi ^2 \zeta_{3}}{9}+\frac{320 \zeta_{5}}{9}
    \approx -291.95399... ~.
\end{align}
It deviates by about 0.1\% from the numerical value reported in
Ref.~\cite{Misiak:2017woa}: $C\approx -292.228$.
In the supplementary material~\cite{progdata} to this paper we provide
computer-readable expressions for $f_0(z)$, $f_1(z)$, $r_{-1}(z)$, $r_0(z)$
and $r_1(z)$.

%- }}}
%- {{{ NNLO:

\subsection{NNLO}

After inserting the three-loop master integrals into the amplitude, we obtain
results for the two-particle cuts at NNLO as series expansions around $x_0=0$
and $x_0=1/5$, with numerical coefficients (see Section~\ref{sec::MIs}).  The
expansions include about 40 terms and we report the $x_0\to0$ expansion in
electronic form in the supplementary material to this paper~\cite{progdata},
where for convenience we keep the colour factors and the electric charges of
the up-type and down-type quarks distinct.

We present in the following the real part of $t_2$ in an expansion up to
$x^2$.  For the colour and charge factors we insert their numerical values.
Furthermore, we provide for all coefficients six significant digits.  Our
results for the operator insertions $Q_1$ and $Q_2$ read
\begin{align}
\mbox{Re}(t_2^{Q_1}) &=
n_l\Bigg\{
-\frac{0.643804}{\ep^2}-\frac{6.31123}{\ep}-27.9137
+x^2 \Bigg[
\frac{1}{\ep}
\Big(2.107 \lm^3+3.16049 \lm^2-27.8263 \lm
\notag \\ &
-11.7523\Big)
-7.37449 \lm^4+3.51166 \lm^3+25.8566 \lm^2-201.543 \lm-247.57\Bigg]\Bigg\}
\notag \\ &
+n_c\Bigg\{
-\frac{0.643804}{\ep^2}-\frac{6.31123}{\ep}-27.9137
+x^2 \Bigg[
\frac{1}{\ep}
\Big(2.107 \lm^3+3.16049 \lm^2-24.6658 \lm
\notag \\ &
-9.61098\Big)
-7.37449 \lm^4+12.9931 \lm^3+54.3011 \lm^2-224.155 \lm-335.398\Bigg]
\Bigg\}
\notag \\ &
+n_b \Bigg\{
-\frac{0.643804}{\ep^2}
-\frac{6.25499}{\ep}
-14.2846
+x^2 \Bigg[
\frac{1}{\ep}
\Big(2.107 \lm^3+3.16049 \lm^2-27.8263 \lm
\notag \\ &
-11.7523\Big)
-5.26749 \lm^4+23.7497 \lm^2-104.437 \lm-132.539\Bigg]
\Bigg\}
\notag \\ &
-\frac{2.0192}{\ep^3}
+\frac{87.3997}{\ep}+256.363
+\frac{8.17904}{\ep^2}
+x \left(\frac{374.314}{\ep}-1497.26 \lm+669.332\right)
\notag \\ &
+x^2 \Bigg[
\frac{1}{\ep^2}
\Big(4.21399 \lm^3+6.32099 \lm^2-55.6525 \lm-23.5046\Big)
+\frac{1}{\ep}
\Big(-13.6955 \lm^4 
\notag \\ &
-36.8724 \lm^3-209.669 \lm^2+1407.45 \lm+233.132\Big)
+27.8123 \lm^5+142.222 \lm^4
\notag \\ &
+402.206 \lm^3-2492.03 \lm^2+7662.75 \lm+8375.85\Bigg]\,,
\\
  \mbox{Re}(t_2^{Q_2}) &=
   n_l \Bigg\{
       \frac{3.86283}{\ep^2 }
    + \frac{37.8674}{\ep} 
    + 167.482 
    +x^2 \Bigg[
    \frac{1}{\ep}
    \Big( 70.5139 
    + 166.958 \lm 
    - 18.963 \lm^2 
    \notag \\ &
    - 12.642 \lm^3 \Big)
          + 1485.42 
    + 1209.26 \lm 
    - 155.14 \lm^2
     - 21.07 \lm^3 + 
      44.2469 \lm^4 \Bigg]
      \Bigg\}
      \notag \\ &
     + n_c \Bigg\{
      \frac{3.86283}{\ep^2}
      +\frac{37.8674}{\ep}
      +167.482
      +x^2 \Bigg[
      \frac{1}{\ep}\Big(
      -12.642 \lm^3
      -18.963 \lm^2
      +147.995 \lm
      \notag \\ &
      +57.6659
      \Big)
      +44.2469 \lm^4-77.9588 \lm^3-325.807 \lm^2+1344.93 \lm+2012.39\Bigg]
      \Bigg\}
      \notag \\ &
      +n_b \Bigg\{
      \frac{3.86283}{\ep^2}
      +\frac{37.53}{\ep}
      +85.7078
      +x^2 \Bigg[
      \frac{1}{\ep}
      \Big(-12.642 \lm^3-18.963 \lm^2+166.958 \lm
      \notag \\ &
      +70.5139\Big)
      +31.6049 \lm^4-142.498 \lm^2+626.621 \lm+795.235\Bigg]
      \Bigg\}
      \notag \\ &
      +\frac{12.1152}{\ep^3}
      -\frac{1.66683}{\ep^2}
      -\frac{148.698}{\ep}+119.784
      +x \left(-\frac{2245.88}{\ep}+8983.53 \lm-4015.99\right)
      \notag \\ &
      +x^2 \Bigg[
      \frac{1}{\ep^2}
      \Big(-25.284 \lm^3-37.9259 \lm^2+333.915 \lm+141.028\Big)
      +\frac{1}{\ep}
      \Big(82.1728 \lm^4
      \notag \\ &
      +50.5679 \lm^3+1002.01 \lm^2-6190.76 \lm-446.854\Big)
      -146.963 \lm^5-355.556 \lm^4
      \notag \\&
      -3798.57 \lm^3+12695.6 \lm^2-30541.1 \lm-32307.3\Bigg] \, ,
\end{align}
where $\lm=\log(x)$ and $n_l=3$ denotes the contribution from
closed massless fermion loops while the $n_c=1$ and $n_b=1$ contributions
arise from closed fermion loops with masses $m_c$ and $m_b$,
respectively. 

In the following we describe the various checks which we performed
on our result. We perform the calculation for generic QCD gauge
parameter $\xi$ which drops out for $t_2^{Q_1}$ and $t_2^{Q_2}$
at the level of the master integrals.
In Ref.~\cite{Bieri:2003ue} analytic expansions for the $z\to0$ limit of the
light-fermion contribution have been computed. We compare the expansion
coefficients of the $x^n\log^k(x)$ terms up to $n=4$ and find agreement with
an accuracy of 15 digits or better. We successfully
compare the heavy-fermion contributions against Refs.~\cite{Boughezal:2007ny,Misiak:2020vlo}.

The authors of Ref.~\cite{Greub:2023msv} have computed the subset of diagrams
contributing to the $b\to s \gamma$ vertex at NNLO where the gluon couples
only to the charm and strange quarks, i.e.\ there is no internal bottom quark
propagator.  Furthermore, all diagrams with fermion, gluon or ghost insertions
in the gluon propagator have been excluded. For the comparison we isolate this
subset of diagrams in our calculation and then compare $t_2$ to the
corresponding expression in the ancillary file of Ref.~\cite{Greub:2023msv}.
We specialize our result to Feynman gauge since this is the choice of
Ref.~\cite{Greub:2023msv} (the set of diagrams considered in
Ref.~\cite{Greub:2023msv} is not gauge invariant).  Using our expansion around
$x_0=1/5$ we observe agreement of more than 10 digits for the terms
proportional to $Q_s$, the charge of the down-type quark, and of about 5
digits for the terms proportional to $Q_c$, the charge of the up-type quark.
According to Ref.~\cite{Greub:2023msv} the program {\tt
  FIESTA}~\cite{Smirnov:2021rhf} has been used to fix parts of the boundary
conditions for the second part. This may explain the reduced accuracy.  Using
our expansion around $x_0=0$ we can also directly compare the coefficients in
front of $x^n \log^k(x)$. As before, we observe about 5 digits agreement
for the $Q_c$ terms and at least 18 digits for the $Q_s$ terms.

In Ref.~\cite{Misiak_et_al} the contributions induced by four-quark operators
are computed via the reverse unitarity method by considering bottom quark
two-point functions with an insertion of $Q_1$ or $Q_2$ and $Q_7$. The three-
and four-loop contributions yield the NLO and NNLO corrections, respectively,
where the latter contains two-, three- and four-particle cuts. In
Ref.~\cite{Misiak_et_al} the two-particle cut contribution is computed.  We
compare our NNLO result for $\hat G_{17}^{(2),2P,Q_7^{\rm tree}}$ and
$\hat G_{27}^{(2),2P,Q_7^{\rm tree}}$ and find agreement at the level of
$10^{-15}$ for $x=1/5$.  Let us stress that the calculations performed in
Ref.~\cite{Misiak_et_al} and the one in the current paper are completely
independent and thus the comparison constitutes an important cross check for
both calculations.

The Ward identity in Eq.~\eqref{eq::WI} holds for renormalized
quantities. Since it holds at NLO, all NNLO contributions obtained by
multiplying the NLO amplitude by a global renormalization factor also
fulfil the Ward identity.  It does not hold for the mass counterterm
contribution and the bare three-loop
amplitude.  However, in the combination of the three-loop corrections
to the form factors and the contributions from the bottom mass
counterterm the relation~\eqref{eq::WI} has to be fulfilled.  For our
calculation we have checked that this is indeed the case. Using the
expansion around $x_0=0$ we observe that the numerical cancellation in
the combination $t_3 + t_2/2$ is of order $10^{-12}$ or smaller
for $x=0.2$ while it reaches the level of $10^{-5}$ at $x=0.4$. At the
latter point an accuracy of at least 12 digits is obtained in case the
expansion around $x_0=1/5$ is used.

%- }}}

%- }}}

%- {{{ Conclusions

\section{\label{sec::conclusions}Conclusions}

In this paper we compute three-loop corrections to the $b\to s \gamma$ vertex
induced by the current-current operators $Q_1$ and $Q_2$. We apply a
semi-analytic method and construct expansions  around $x_0=0$, $x_0=1/10$ and
$x_0=1/5$ including about 40 terms,
which covers the physically relevant values of the mass ratio in different
mass schemes.
Furthermore, we provide for the first time analytic results for the two-loop
contributions.

The current predictions for the $B \to X_s \gamma$ branching ratio at NNLO
employ only an interpolation to estimate the $m_c$ dependence in the
interference term between $Q_1,Q_2$ and $Q_7$. Such interpolation is
responsible for 3\% out of the 5\%
$(=\sqrt{(3\%)^2+(3\%)^2+(2.5\%)^2})$ theoretical uncertainty.  Our result for
the $b\to s \gamma
$ vertex at NNLO allows to calculate the two-particle cut contribution to
the $B \to X_s
\gamma$ branching ratio (stemming from current-current operators) without
such an interpolation. Therefore it constitutes an important building block
towards the reduction of the theoretical error in $B \to X_s
\gamma$, once the missing three- and four-particle cut contributions are
available as well.

%- }}}

\section*{Acknowledgements}  

We thank Mikolaj Misiak for useful comments to the manuscript.
This research was supported by the Deutsche Forschungsgemeinschaft
(DFG, German Research Foundation) under grant 396021762 --- TRR 257
``Particle Physics Phenomenology after the Higgs Discovery'' and has
received funding from the European Research Council (ERC) under the
European Union’s Horizon 2020 research and innovation programme grant
agreement 101019620 (ERC Advanced Grant TOPUP).  The work of M.F. was
supported by the European Union’s Horizon 2020 research and innovation
program under the Marie Sk\l{}odowska-Curie grant agreement
No.~101065445 - PHOBIDE.  We have used the program {\tt
  FeynGame}~\cite{Harlander:2020cyh} to draw the Feynman diagrams.

%\newpage
\appendix

%- {{{ Details of the two-loop master integral calculation:

\section{\label{app::2loop}Details of the two-loop master integral calculation}

In this appendix we present  additional details regarding the 
calculation of the master integrals at two loops, in particular 
about the analytic continuation.

As discussed in Section~\ref{sec::MIs}, at two loops we solve the
master integrals using the method of differential equations and fix
the boundary conditions in the limit $m_c \to \infty$ via a large mass
expansion.  To match the boundary conditions it is
convenient to first solve the differential equations in the variable
$y = m_b/m_c$ since the expansions for $y \to 0$ are easier to
compute.  It is therefore necessary at some point to perform an
analytic continuation to the variable $x = m_c/m_b$.  We find it
beneficial to use the argument $y$ for the master integrals and only
apply the appropriate variable change once the physical amplitude is
assembled. In the following we explain how the analytic continuation
is performed.

We express the master integrals in terms of iterated integrals with
letters from the set given in Eq.~\eqref{eq::alphabet}.  In the end we
want to evaluate the iterated integrals for values $m_c/m_b < 1$,
i.e.\ $y>1$.  As we can see in Eq.~\eqref{eq::alphabet} the iterated
integrals therefore require an analytic continuation since the letters
develop poles at $y=1$ and $y=2$.  For iterated integrals which only
contain the first three letters, i.e.\ harmonic polylogarithms, the
analytic continuation to $x=1/y$ is well known; we use {\tt
  HarmonicSums} to do this step.  For the integrals which contain
non-standard letters we do the analytic continuation with the help of
the differential equations.  We start by taking the derivative of the
iterated integrals with respect to $y$, which leads to iterated
integrals of lower weight, then do the variable change $x = 1/y$ and
integrate over $x$.  The differential equations with iterated
integrals of weight~1 can then be integrated trivially since they
reduce to integrals over algebraic functions which can be performed by
using the definition of iterated integrals and possibly
integration-by-parts identities to reduce higher powers of the
letters. For the contributions with higher weight one can proceed
iteratively, since the derivative of an iterated integral of weight
$n$ with respect to its argument only depends on iterated integrals of
weight $n-1$.  After the change of variables we therefore have to
perform a single integral over expressions of algebraic functions
possibly multiplying iterated integrals of lower weight. These
integrals can again be performed using the definition of iterated
integrals and integration-by-parts identities.  However, in order to
find the analytic continuation of the iterated integrals in this way
we have to fix the integration constants.

It is convenient to fix this integration constants at $y=2$ since
after the shift $y\to 2\tilde{y}$ the non-standard letters in
Eq.~(\ref{eq::alphabet}) transform to $1/(1+\tilde{y})$,
$1/(1-\tilde{y}),\sqrt{1-\tilde{y}^2}$.  The square root valued letter
can further be rationalized which subsequently leads to cyclotomic
harmonic polylogarithms~\cite{Ablinger:2011te} for which the constants
at argument $\tilde{y}=1$ are known.

Note that it is crucial to consider the analytic continuation of
the whole amplitude and not only a single iterated integral.  The
individual integrals depend on various constants for which we do not
have a replacement in terms of known transcendental numbers.  However,
in the amplitude these cancel out and we are left with well known
constants.  We also observe that the analytic continuation of single
master integrals is much more involved than the one of the physical
amplitude.

Finally, we notice that the analytic continuation of the HPLs
completely decouples from the one of the generalized iterated
integrals.  This can be seen by introducing $y = 1/x + i
\delta^{(')}$, where $\delta$ is used for the harmonic polylogarithms
and $\delta^{'}$ for the generalized iterated integrals. Both quantities
are infinitesimally small. After
performing the analytic continuation we see that the expression is
independent of $\delta$ while we have to choose $\delta^{'}$ as positive in
order to identify the correct analytic continuation for the
generalized iterated integrals.  This effect is not necessarily
expected.  In other words: From the observation that the analytic
continuation of the HPLs is independent from the one of the
``non-standard'' iterated integrals we conclude that they ``decouple''
and a separate variable transformation can be applied to the latter.
Note that this discussion only affects the imaginary parts of
$t_2^{Q_1}$ and $t_2^{Q_2}$ and is thus not relevant for the results
presented in the main part of this paper.

%- }}}
%- {{{ Results for a and b:

\section{\label{app::ab}Analytic results for $a(z)$ and $b(z)$}

The functions $a(z)$ and $b(z)$ defined in Ref.~\cite{Buras:2002tp} can be obtained as the
contributions proportional to $Q_u$ and $Q_d$ with an additional subtraction
to fulfill the normalization $a(0)=b(0)=0$.  In the conventions of
Ref.~\cite{Buras:2002tp} we find
\begin{align} 
  a(z) &= 
  \frac{16 w}{3 (1+w)^2}
  +\frac{8 w \big(5+w^2\big) H_0(w)}{9 (1+w)^3}
  -\frac{4 w \big(-1+w+w^2\big) H_0(w)^2}{9 (1+w)^4}
  +\frac{4 w H_0(w)^3}{9 (1+w)^2}
  \nonumber \\ &
  -\frac{32 w \big(2+3 w+2 w^2\big) H_0(x)^3}{27 (1+w)^4}
  -\frac{8 \big(1+4 w+w^2\big) H_{-1}(w)}{9 (1+w)^2}
  +\frac{16 w^2 (1+2 w) H_{0,-1}(w)}{9 (1+w)^4}
  \nonumber \\ &
  -\frac{16 w \big(1-\sqrt{w}+w\big)\big(1+\sqrt{w}+w\big) H_{1,0}(x)}{9 (1+w)^4}
  +\frac{16 w \big(1+w+w^2\big) H_{-1,0}(w)}{9 (1+w)^4}
  \nonumber \\ &
  +\frac{16 w \big(1-\sqrt{w}+w\big)\big(1+\sqrt{w}+w\big) H_{-1,0}(x)}{9 (1+w)^4}
  -\frac{32 w \big(1+w+w^2\big) H_{-1,-1}(w)}{9 (1+w)^4}
  \nonumber \\ &
  -\frac{16 w \big(1+w^2\big) H_{0,0,-1}(w)}{9 (1+w)^4}
  -\frac{32 w \big(2+3 w+2 w^2\big) H_{0,1,0}(x)}{9 (1+w)^4}
  -\frac{16 w \zeta_3}{3 (1+w)^2}
  \nonumber \\ &
  +\frac{32 w \big(2+3 w+2 w^2\big) H_{0,-1,0}(x)}{9 (1+w)^4}
  -\frac{32 w \big(1+3 w+w^2\big) H_{-1,0,0}(w)}{9 (1+w)^4}
  \nonumber \\ &
  +\pi ^2 
  \biggl(
        -\frac{8 w \big(2+w+3 w^2\big)}{27 (1+w)^4}
        -\frac{4 w \big(3+8 w+3 w^2\big) H_0(w)}{9 (1+w)^4}
        \nonumber \\ &
        +\frac{16 w \big(5+12 w+5 w^2\big) H_{-1}(w)}{27 (1+w)^4}
  \biggr)
  + {\rm i} \pi 
  \biggl(
        \frac{8 w \big(4-w+w^2\big)}{9 (1+w)^3}
        -\frac{8 \pi ^2 w \big(1+3 w+w^2\big)}{27 (1+w)^4}
        \nonumber \\ &
        +\frac{8 w \big(1+w^2\big) H_0(w)}{9 (1+w)^4}
        +\frac{8 w \big(1+3 w+w^2\big) H_0(w)^2}{9 (1+w)^4}
        -\frac{32 w \big(1+3 w+w^2\big) H_{-1,0}(w)}{9 (1+w)^4}
  \biggr)
  ~, \\
  b(z) &=
  \frac{32 w \big(7+17 w+7 w^2\big)}{81 (1+w)^4}
  -\frac{4 \pi ^2 w}{81 (1+w)^6} \big(3+30 w+52 w^2+42 w^3+9 w^4-12 \sqrt{w}
  \nonumber \\ &
  -36 w^{3/2}-36 w^{5/2}-12 w^{7/2}\big)
  +\frac{8 w \big(9+24 w-2 w^2-4 w^3+w^4\big) H_0(w)}{81 (1+w)^5}
  \nonumber \\ &
  -\frac{4 w^2 \big(-1+2 w+3 w^2+2 w^3\big) H_0(w)^2}{9 (1+w)^6}
  -\frac{8 \big(1+4 w+18 w^2+4 w^3+w^4\big) H_{-1}(w)}{81 (1+w)^4}
  \nonumber \\ &
  +\frac{16 w^2 \big(3+14 w+15 w^2+6 w^3\big) H_{0,-1}(w)}{27 (1+w)^6}
  +\frac{16 w \big(3+9 w+14 w^2+9 w^3+3 w^4\big) H_{-1,0}(w)}{27 (1+w)^6}
  \nonumber \\ &
  -\frac{16 w \big(1-\sqrt{w}+w\big)^2 \big(3+8 w+3 w^2+4 \sqrt{w}+4 w^{3/2}\big) H_{1,0}(x)}{27 (1+w)^6}
  \nonumber \\ &
  +\frac{16 w \big(1+\sqrt{w}+w\big)^2 \big(3+8 w+3 w^2-4 \sqrt{w}-4 w^{3/2}\big) H_{-1,0}(x)}{27 (1+w)^6}
  \nonumber \\ &
  -\frac{32 w \big(3+9 w+14 w^2+9 w^3+3 w^4\big) H_{-1,-1}(w)}{27 (1+w)^6}
  \nonumber \\ &
  + {\rm i} \pi 
  \biggl(
    \frac{8 w \big(9+18 w-8 w^2-4 w^3+w^4\big)}{81 (1+w)^5}
    +\frac{16 w^2 \big(3+4 w+3 w^2\big) H_0(w)}{27 (1+w)^6}
  \biggr)
  ~.
\end{align}
The variables $x$ and $w$ are defined in Eqs.~(\ref{eq::x}) and~(\ref{eq::w}),
respectivley.
Note that the separation in real and imaginary parts is correct 
below the two-particle threshold $z<1/4$.

From our expressions we can find analytic expressions for the constants in
Ref.~\cite{Buras:2002tp}. They agree with the results derived in
Ref.~\cite{Czakon:2015exa} by a direct integration of the three-fold
integral representation.  They are given by
\begin{align}
  a(1) &= 
  \frac{16}{3}
  +\frac{164 \pi ^2}{405}
  -\frac{8}{45} \psi ^{(1)}\left(\frac{1}{6}\right)
  + \frac{\pi}{\sqrt{3}}  
  \biggl[
        -\frac{20}{9} 
        -\frac{64}{135} \pi^2
        +\frac{32}{135} \psi ^{(1)}\left(\frac{1}{6}\right)
  \biggr]
  \nonumber \\ &
  -\frac{16 \zeta_3}{9}
  + \frac{4}{9} {\rm i} \pi
  ~, \\ 
  b(1) &= 
  \frac{320}{81}
  +\frac{632 \pi ^2}{1215}
  -\frac{4}{3} \frac{\pi}{\sqrt{3}}
  -\frac{8}{45} \psi ^{(1)}\left(\frac{1}{6}\right)
  + \frac{4}{81} {\rm i} \pi
  ~, \\
  X_b &= 
  -\frac{9}{8}
  -\frac{\pi ^2}{5}
  +\frac{1}{10} \psi ^{(1)}\left(\frac{1}{6}\right)
  -\frac{2 \zeta_3}{3}
  ~.
\end{align}
The results in this Appendix together with the expressions
given in Eq.~(3.1) of Ref.~\cite{Buras:2002tp} provide
analytic results also for all NLO penguin contributions.

In Refs.~\cite{deBoer:2017way,Asatrian:2019kbk} the two-loop form factors with an off-shell
photon have been calculated.  Obtaining the on-shell limit from the
representation in terms of Goncharov multiple polylogarithms is, however, a
non-trivial task, since many letters become singular in this limit.

%- }}}

%\newpage

%%%\bibliographystyle{JHEP} 
%%%\footnotesize
%%%\bibliography{BIB}

\end{document}